\documentclass[sigconf]{acmart}

\AtBeginDocument{%
  \providecommand\BibTeX{{%
    \normalfont B\kern-0.5em{\scshape i\kern-0.25em b}\kern-0.8em\TeX}}}

\copyrightyear{2022}
\acmYear{2022}
\setcopyright{acmcopyright}

\acmConference[CSCW'22 Companion] {Companion Computer Supported Cooperative Work and Social Computing}{November 08--22, 2022}{Virtual Event, Taiwan}

\acmBooktitle{Companion Computer Supported Cooperative Work and Social Computing (CSCW'22 Companion), November 08--22, 2022, Virtual Event, Taiwan}
\acmPrice{15.00}
\acmDOI{10.1145/3500868.3559706}
\acmISBN{978-1-4503-9190-0/22/11}

\usepackage{tabularx}
\usepackage{hyperref}

\usepackage{subcaption}
\usepackage{enumitem}

\usepackage{float} 
\usepackage{soul,color}
\usepackage[utf8]{inputenc}

\usepackage{chngcntr}
\usepackage{xcolor}
\definecolor{editCol}{rgb}{0.0, 0.0, 0.0}

\usepackage{pbox}
\usepackage{multirow}
\usepackage[colorinlistoftodos]{todonotes}
\usepackage{array}
\usepackage{makecell}
\usepackage{afterpage} 
\usepackage[title]{appendix}
\usepackage{subcaption,graphicx}
\usepackage{ragged2e}

\newcolumntype{L}[1]{>{\raggedright\let\newline\\\arraybackslash\hspace{0pt}}m{#1}}
\newcolumntype{C}[1]{>{\centering\let\newline\\\arraybackslash\hspace{0pt}}m{#1}}
\newcolumntype{R}[1]{>{\raggedleft\let\newline\\\arraybackslash\hspace{0pt}}m{#1}}
\begin{document}

\title[Community Oversight for Privacy and Security]{CO-oPS: A Mobile App for Community Oversight of Privacy and Security}


\author{Mamtaj Akter}
\orcid{0000-0002-5692-9252}
\affiliation{%
  \institution{Vanderbilt University}
  \city{Nashville}
  \state{Tennessee}
  \postcode{37235}
  \country{USA}
}
\email{Mamtaj.Akter@vanderbilt.edu}

\author{Leena Alghamdi}
\orcid{0000-0003-2102-9155}
\affiliation{%
  \institution{University of Central Florida}
  \city{Orlando}
  \state{Florida}
  \postcode{32826}
    \country{USA}
}
\email{Leenaalghamdi@knights.ucf.edu}

\author{Dylan Gillespie}
\orcid{0000-0003-1091-8029}
\affiliation{%
  \institution{University of Central Florida}
  \city{Orlando}
  \state{Florida}
  \postcode{32826}
    \country{USA}
}
\email{Dgillespie00@knights.ucf.edu}

\author{Nazmus Sakib Miazi}
\orcid{0000-0002-9778-4678}
\affiliation{%
  \institution{Northeastern University}
  \city{Boston}
  \state{Massachusetts}
  \postcode{02115}
    \country{USA}
}
\email{M.Miazi@northeastern.edu}

\author{Jess Kropczynski}
\orcid{0000-0002-7458-6003}
\affiliation{%
  \institution{University of Cincinnati}
  \streetaddress{P.O. Box 1234}
  \city{Cincinnati}
  \state{OH}
  \postcode{45221}
    \country{USA}
}
\email{Jess.Kropczynski@uc.edu}

\author{Heather Lipford}
\orcid{0000-0002-5261-0148}
\affiliation{%
  \institution{University of North Carolina, Charlotte}
  \city{Charlotte}
  \state{North Carolina}
  \postcode{28223}
     \country{USA}
}
\email{Heather.Lipford@uncc.edu}

\author{Pamela J. Wisniewski}
\orcid{0000-0002-6223-1029}
\affiliation{%
  \institution{Vanderbilt University}
  \city{Nashville}
  \state{Tennessee}
  \postcode{37235}
    \country{USA}
}
\email{Pamela.Wisniewski@vanderbilt.edu}

\renewcommand{\shortauthors}{Mamtaj Akter et al.}

\begin{abstract}
Smartphone users install numerous mobile apps that require access to different information from their devices. Much of this information is very sensitive, and users often struggle to manage these accesses due to their lack of tech expertise and knowledge regarding mobile privacy. Thus, they often seek help from others to make decisions regarding their mobile privacy and security. We embedded these social processes in a mobile app titled "CO-oPS'' ("Community Oversight for Privacy and Security"). CO-oPS allows trusted community members to review one another's apps installed and permissions granted to those apps. Community members can provide feedback to one another regarding their privacy behaviors. Users are also allowed to hide some of their mobile apps that they do not like others to see, ensuring their personal privacy.
\end{abstract}

\ccsdesc[500]{Security and privacy~Social aspects of security and privacy}

\keywords{Community Oversight; Mobile Privacy; Online Safety; Android phone; Mobile Apps; App Permissions}

\maketitle
\section{Introduction}
85\% of US citizens own smartphones ~\cite{nw_demographics_nodate} and 77\% of them reported that they downloaded and installed mobile applications ("apps") on their smartphones ~\cite{noauthor_majority_2015}. These mobile apps often require access to users' sensitive information, like contact data, emails, location, calendars, and even browser history ~\cite{nw_mobile_2015}. Although most of these apps request users' permission before accessing any information or resources, many apps secretly gathered users' system resources (e.g., camera, GPS) and private information (e.g., contacts list, text messages, emails) without users' consent ~\cite{reardon_50_2019, calciati_automatically_2020}. Therefore, most smartphone users are concerned about their mobile information privacy as they are not aware of how these mobile apps are using these resources ~\cite{davis_perceived_1989}. This lack of transparency and users' lack of privacy and security knowledge cause them to seek advice and guidance from their close ones for making their digital privacy and security choices ~\cite{dourish_security_2004}. People also often learn about privacy and security from others in their social network, and this indirect learning eventually influences them to change their privacy behavior ~\cite{schechter_learning_2015, felt_android_2012}. 

Therefore, many researchers have acknowledged the importance of these social processes for managing individual, and collective digital privacy and security ~\cite{das_effect_2014, mendel_susceptibility_2017, rader_identifying_2015}. Some other recent studies proposed mechanisms that allowed trusted members of a community to help one another in making their digital privacy and security decisions ~\cite{aljallad_designing_2019, chouhan_co-designing_2019}. For example, Chouhan et al. proposed a mechanism titled Community Oversight for Privacy and Security ("CO-oPS") that allowed a group of trusted members of a community to help one another manage their mobile privacy and security, utilizing the concepts of individual participation, transparency, trust, and awareness. We converted this Community Oversight mechanism into a mobile app titled CO-oPS ~\cite{chouhan_co-designing_2019}. CO-oPS allows individuals to review the apps installed and permissions granted on their community members' phones. It also lets users provide direct feedback to one another about their privacy and security behavior.

\section{Background}


The proliferation of smartphone devices and the usage of mobile applications caused mobile phone users be over-exposed to the app permission requests and therefore they often overlook the permission prompts ~\cite{till_characterization_2019}. Mobile app users often do not fully understand what these mobile app permissions do and what are they used for and they do not even know where their data are being sent through accepting these permissions ~\cite{ferreira_securacy_2015, alsoubai_permission_2022, ghaiumy_disclose_2021}. Recent privacy research ~\cite{calciati_automatically_2020, reardon_50_2019} also reported that many third party mobile apps automatically grant some permissions and accessed users' sensitive information which the users never explicitly acknowledged ~\cite{reardon_50_2019}. 

Due to the lack of privacy awareness and transparency among these app permissions, technology users often seek advice and guidance from their loved ones to make decisions regarding privacy and security ~\cite{dourish_security_2004, 10.1145/3479540}. People also learn from their social network and eventually get influenced to change their privacy behavior ~\cite{schechter_learning_2015,felt_android_2012}. Hence, many network privacy researchers emphasized the importance of social collaboration ~\cite{das_effect_2014, badillo2020towards, rader_identifying_2015} in managing privacy and security.  With a goal to implement a technological solution to provide this social collaboration ability, we developed a novel mobile application titled CO-oPS (Community Oversight for Privacy and Security). The purpose of this app is to help mobile users mitigate their privacy awareness and knowledge gap by allowing them interact with their community and work together to keep their information safe from third party mobile apps. The next section discusses the overview and functionality of the features of our proposed mobile privacy app CO-oPS.

\begin{figure}
\begin{subfigure}[t]{.30\textwidth}
\centering
  \includegraphics[width=\columnwidth]{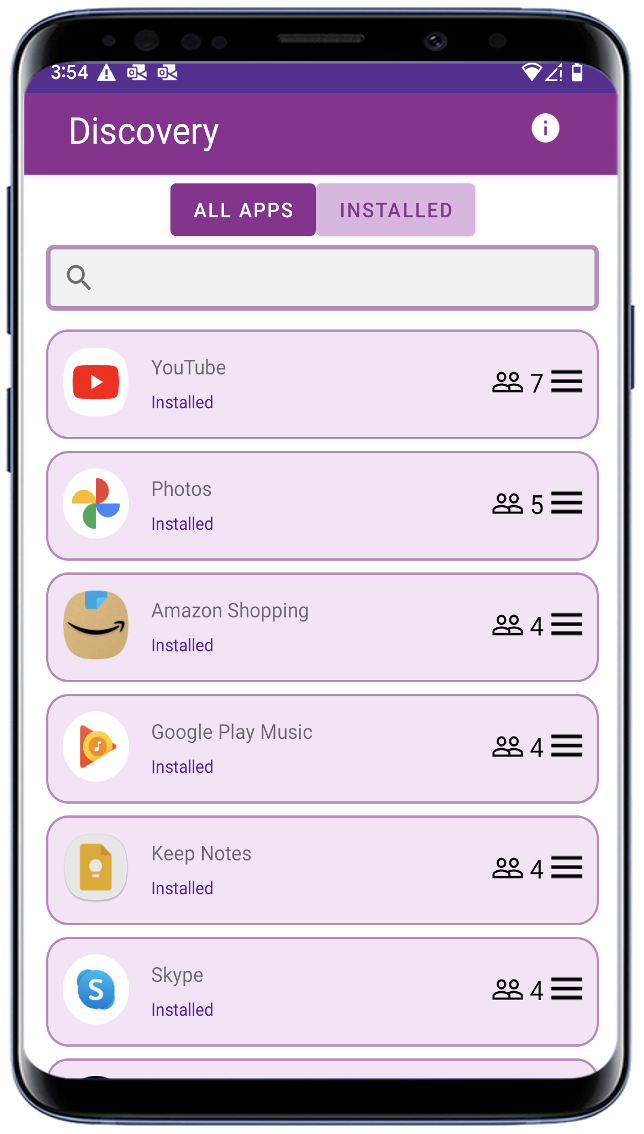}
 \Description[A CO-oPS feature]{A feature that shows Community Apps}
 \end{subfigure}
\caption{Community Apps}
    \label{fig:figure1}
\end{figure}

\begin{figure}
\begin{subfigure}[t]{.30\textwidth}
\centering
  \includegraphics[width=\columnwidth]{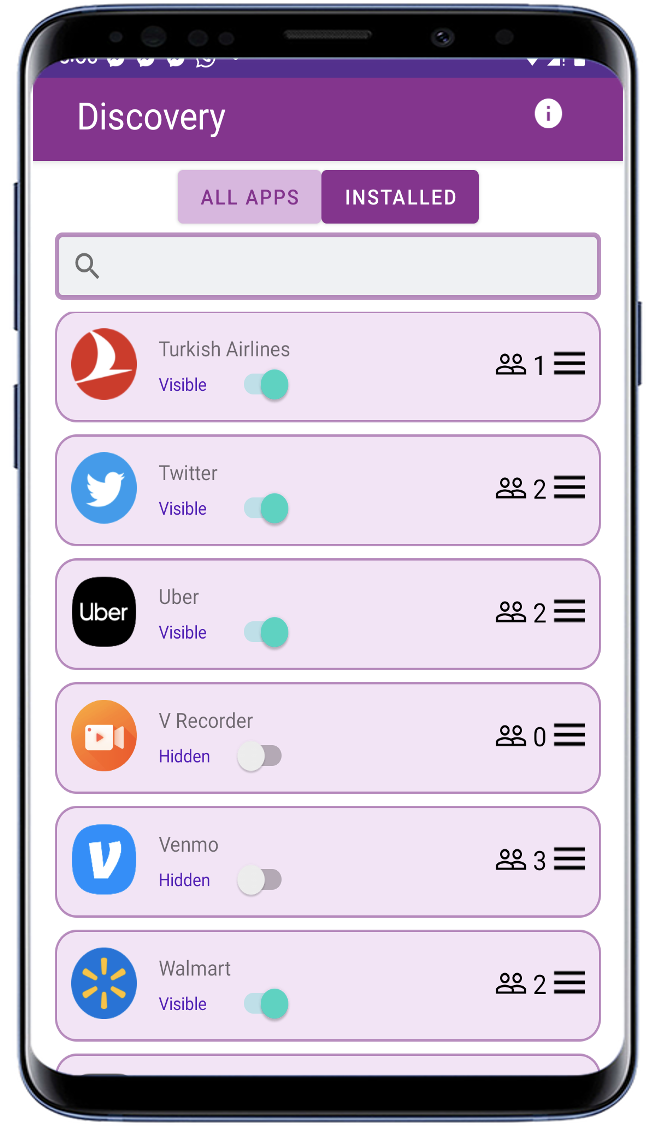}
 \Description[A CO-oPS feature]{A feature that shows User's own apps}
 \end{subfigure}
\caption{Own Apps}
    \label{fig:figure2}
\end{figure}

\begin{figure}
\begin{subfigure}[t]{.30\textwidth}
\centering
  \includegraphics[width=\columnwidth]{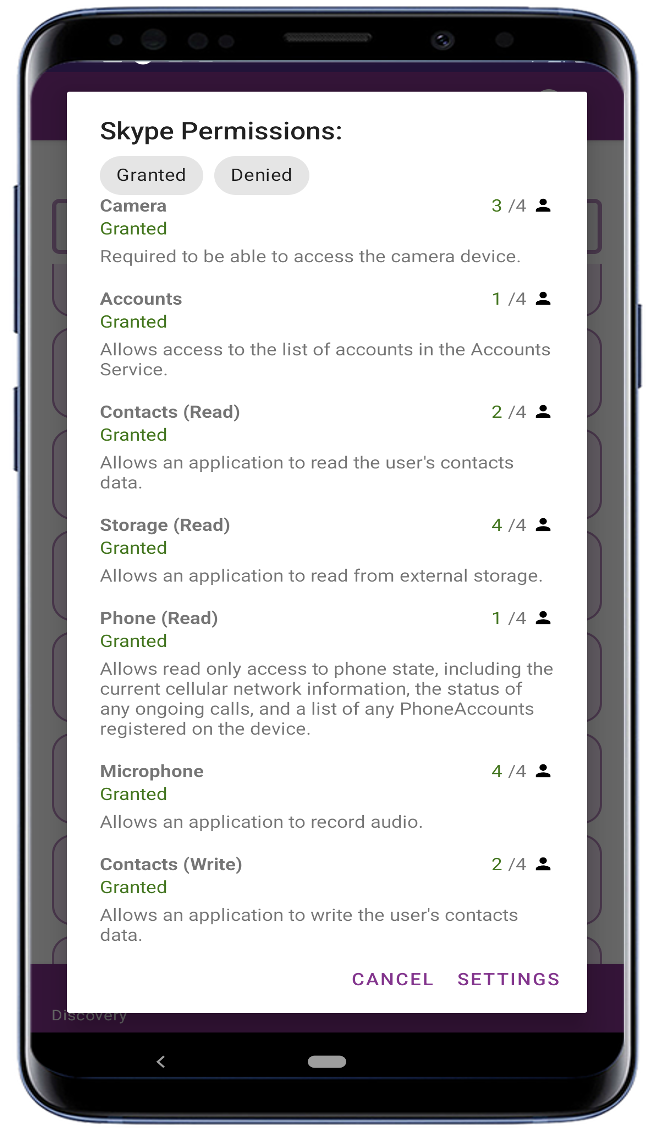}
 \Description[A CO-oPS feature]{A feature that shows App Permissions}
 \end{subfigure}
\caption{App Permissions}
    \label{fig:figure3}
\end{figure}


\section{CO-oPS App Design}
CO-oPS allows all community members to review the apps installed on one another's phones. It also allows checking what permissions are granted or denied to the installed apps. So, CO-oPS does not just let users monitor other community members' apps installed; it enables them to watch whether these apps access any sensitive data (e.g., contacts, emails, photos, location, browser history) from their phones. It also allows users to hide any of their own apps that they are not comfortable sharing with others, supporting their personal privacy. We developed this app based on the design suggestions made in our previous study by Chouhan et al. \cite{chouhan_co-designing_2019}. They proposed a novel mechanism for users to interact with people they trust to help one another make digital privacy and security decisions regarding mobile app permissions. Their participatory design study provided some design suggestions to translate this framework into mobile app features. We leveraged those feature suggestions and implemented a full-functioning mobile app in this work. CO-oPS app includes six main features: 1) Community Apps, 2) Own Apps, 3) App Permissions, 4) Community Members, 5) Individual Apps, and 6) Community Feed.

\textit{ - Community Apps (Figure-\ref{fig:figure1}):} Under the Discovery tab, the All Apps section presents the list of all apps that are installed on all community members' phones. The icon and name of the app is displayed, along with the word "Installed" if the user has that app on their phone. On the right side of the screen is a number indicating how many members of the community have the app installed, along with a three line menu icon that will open an app permissions dialogue box. The All Apps section can be used to find apps that others in the community are using and find potentially concerning apps that a user may want to discuss with the other members.

\textit{ - Own Apps (Figure-\ref{fig:figure2}): } The Installed section displays only apps that the user has installed on their own device. Alongside the icon and name of each app is a switch that allows users to toggle the visibility of each app between "Visible" and "Hidden". Apps on your device that are changed to hidden will not be viewable to other members of the community. Information shown on the right side of the screen for each app is the same as in the All Apps section.

\textit{ - App Permissions (Figure-\ref{fig:figure3}):} The App Permissions section appears as a dialogue box and shows how many community members have granted or denied a permission for a specific app. For each permission an app requests there will be a description of what it accesses and a readout of how many community members out of the total number have granted or denied a permission. If members of the community have granted the permission the word "Granted" will be displayed in green text. If they have denied the permission the word "Denied" will be displayed in red text. The permissions can be filtered by Granted or Denied by tapping the labeled buttons at the top of the screen. This section can also bring users directly to their phones app permissions settings menu by tapping the settings button at the bottom of the page. This section can be used to compare what permissions the user has granted with their community and see if they are in consensus about what permissions to grant to certain apps.  If they are not this may lead to a discussion on what permissions ought to be granted.

\begin{figure}
\begin{subfigure}[t]{.30\textwidth}
\centering
  \includegraphics[width=\columnwidth]{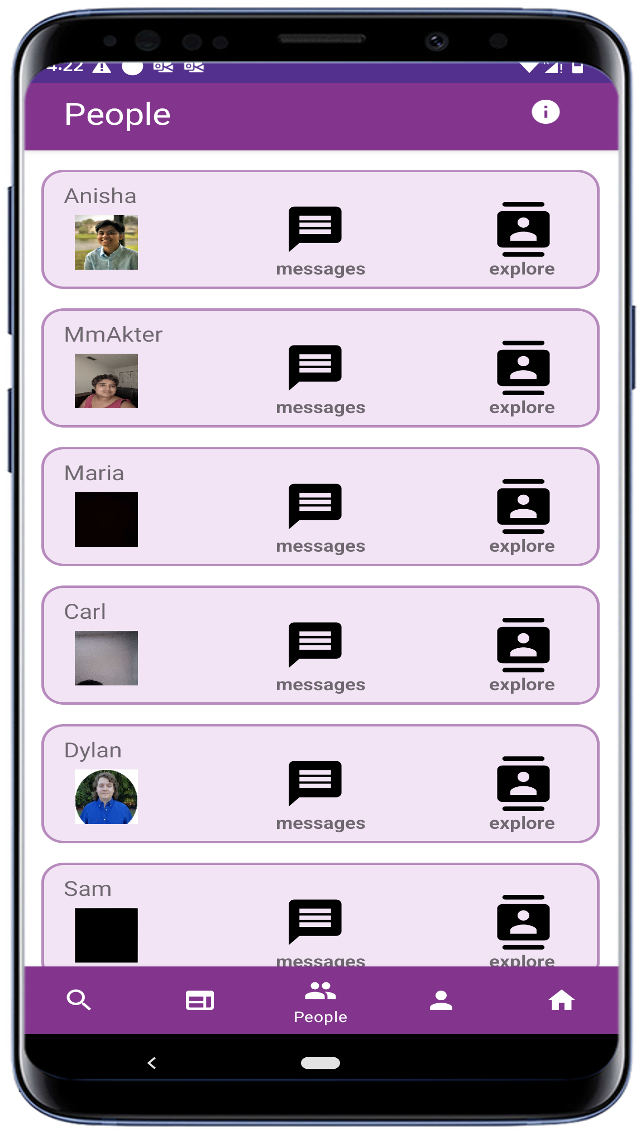}
 \Description[A CO-oPS feature]{A feature that shows list of the community members}
 \end{subfigure}
\caption{Community Members}
    \label{fig:figure4}
\end{figure}

\begin{figure}
\begin{subfigure}[t]{.30\textwidth}
\centering
  \includegraphics[width=\columnwidth]{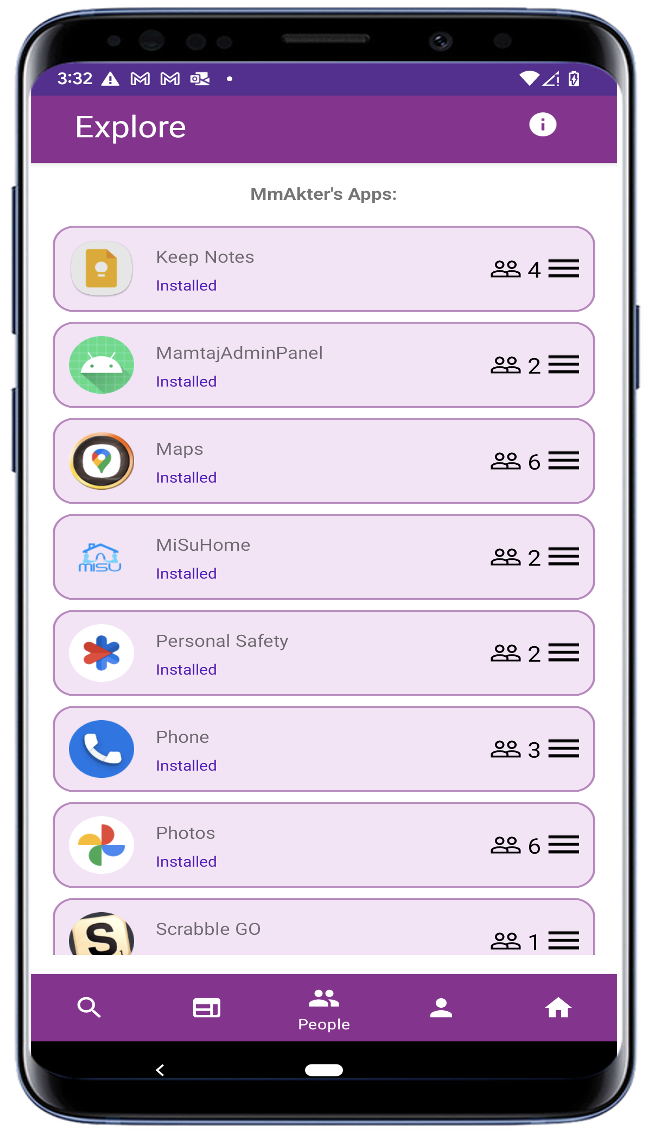}
 \Description[A CO-oPS feature]{A feature that shows each of the community member's apps}
 \end{subfigure}
\caption{Individual's Apps}
    \label{fig:figure5}
\end{figure}

\begin{figure}
\begin{subfigure}[t]{.30\textwidth}
\centering
  \includegraphics[width=\columnwidth]{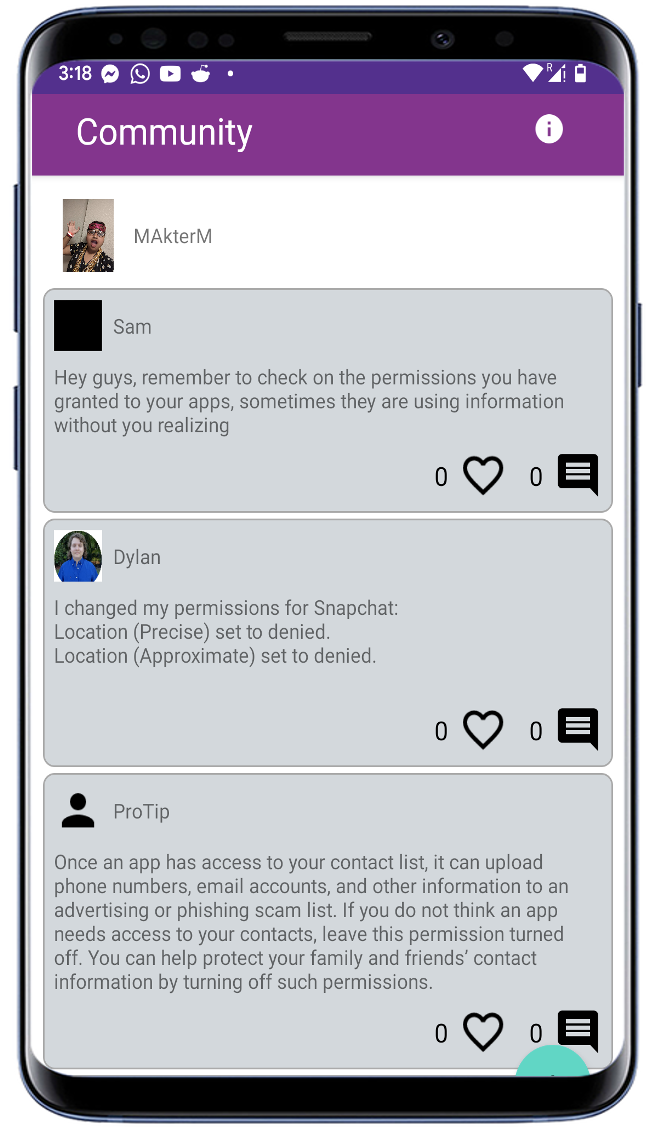}
 \Description[A CO-oPS feature]{A feature that shows community posts}
 \end{subfigure}
\caption{Community Feed}
    \label{fig:figure6}
\end{figure}


\textit{ - Community Members (Figure-\ref{fig:figure4}): } Information relating a specific community member can be found within the People tab. Here a profile picture for each community member is displayed along with "messages" and "explore" buttons. The messages button next to each community members will take users to a page where they can message that specific member to discuss privacy issues or give feedback about suspicious apps or app permissions. The explore button allows users to view the apps that a specific community member has installed.

\textit{ - Individual Apps (Figure-\ref{fig:figure5}): } Within the Explore section a user can see the apps installed on an individual community member's device. This section is laid out like the All Apps section on the discovery page except it only lists apps that the selected community member has installed. Tapping the three line menu icon will take the user to the app permissions page for that app. The explore feature could be used to locate a suspicious app that a community member have installed so that the user to could then message that specific community member to provide guidance about their apps.

\textit{ - Community Feed (Figure-\ref{fig:figure6}): } The community feed functions like a forum where community members can make posts and others can like those posts and reply to them. This feed can be used to initiate discussion with all other community members about specific apps, permissions or any general strategies to keep the community safe. The community feed also provides weekly pro-tips from the CO-oPS app to help community members remain aware with regards to privacy, security installing new apps, and granting permissions.
 
\begin{figure}[t]
\centering
  \includegraphics[width=\columnwidth]{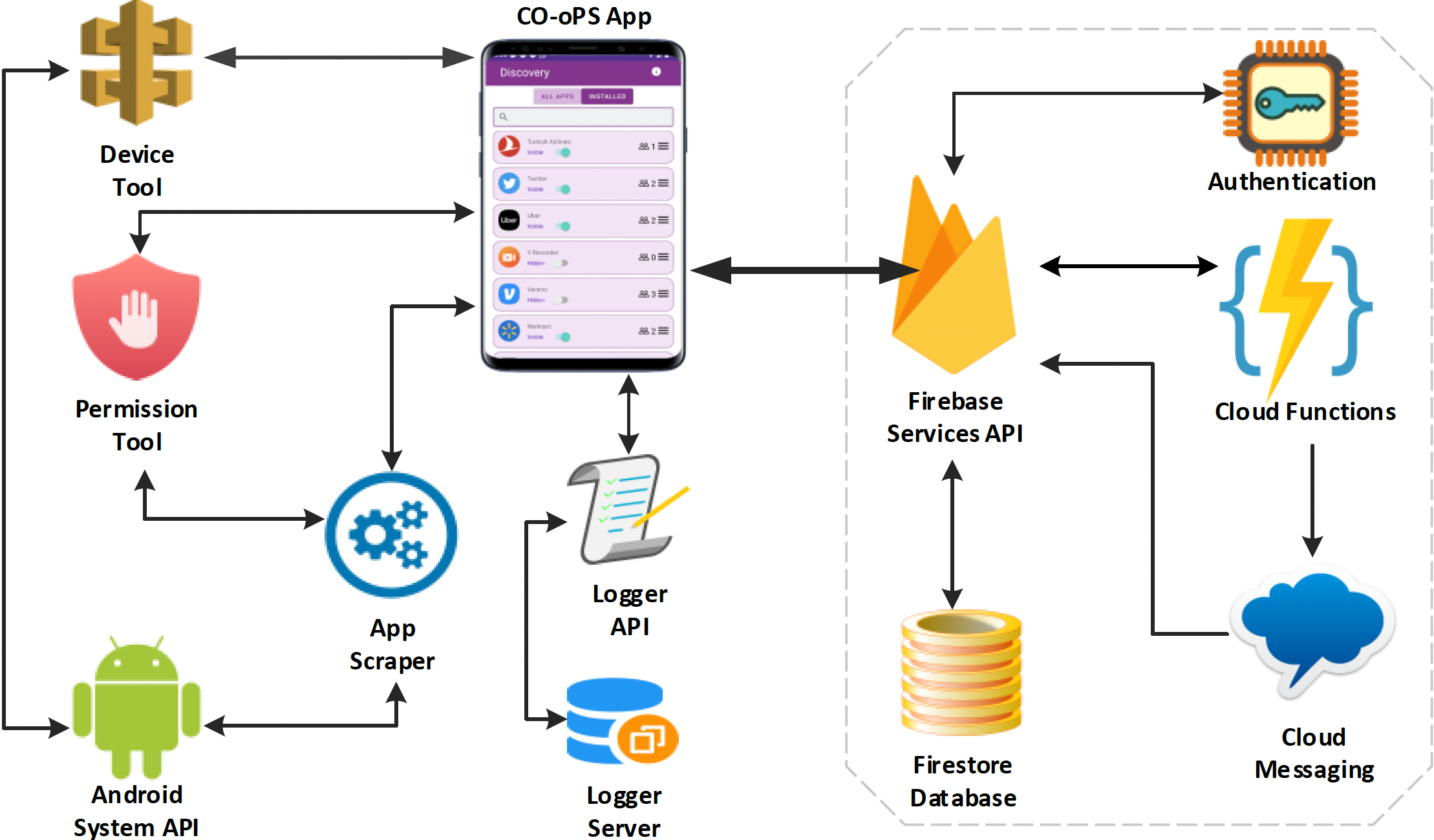}
 \Description[CO-oPS System Diagram]{An image that shows the System Architecture of the CO-oPS App}
\caption{CO-oPS System Architecture}
    \label{fig:figure7}
\end{figure}


\section{System Architecture}
The CO-oPS app is developed with an Android native front-end and a hybrid back-end consisting of Firebase cloud services and NodeJS servers. It uses three main back-end APIs (Figure-\ref{fig:figure7}) for its user interactions: 1) AppScraper, 2) DeviceTool, and 3) PermissionTool. AppScraper is a utility designed to execute background threads to access local on-device data using Android systems API. This tool collects app metadata on the device, compares it against the remote catalog of apps, and sends any relevant changes to Firebase. Through the DeviceTool API, the device ID, SIM serial number, and Android ID are fetched. The PermissionTool API serves to convey the permissions associated with a given app. In addition to these three, the CO-oPS app uses a custom-built authentication API that employs Firebase and DeviceTool API to fit the unique need to form CO-oPS groups. Also, the app uses an API with a NodeJS back-end to store the anonymized logs of the app usage. Lastly, this app consists of a push notification API managed by Firebase Cloud Functions to maintain a seamless and effective group interaction.

\section{Limitations and Future Work}
 While the CO-oPS app provides many important benefits, e.,g., a collaborative platform to co-manage mobile privacy, and personal privacy by hiding apps, it still has some drawbacks that we intend to mitigate in the future. One of the most significant limitations of this app is it does not provide any suggestions or recommendations regarding the app permissions. When users are less tech-savvy or have less knowledge and awareness about mobile app permissions, they might not know which permissions are safe or dangerous. Although this app currently provides weekly pro tips to educate users, in a future version of this app, we intend to implement push notifications to alert the users when any of their apps acquire dangerous permissions (e.g., account, location, contacts). Additionally, CO-oPS app falls short in usability. The user might want to review their apps by the permission names. For example, they might want to view the group of apps that have a specific permissions granted. For example, a user may want to view the list of their apps that have the precise location permission granted. The primary purpose of this app is to help communities in securing their information from third-party apps, and so it is crucial to redesign this app such that users can view and group their apps by the permissions granted. Additionally this app may not be applicable among families with hierarchical tensions (e.g., parents and teens) as it allows privacy in their app usage and equal power in co-monitoring \cite{10.1145/3512904}. Future iterations of this app need to consider such cases to allow family members to help one another manage their mobile privacy. We also can examine whether this app can help parents and teens be influenced by one another's app usage and permission decisions to change their online safety \cite{10.1007/978-3-030-77392-2_17} and privacy behaviors. Lastly, we intend to launch a longitudinal field study where groups of people (friends, families, communities) can try different features of this app and give their feedback on a weekly basis.   

\section{Conclusion}
Our CO-oPS app represents a shift from an individual's effort to a collaborative relationship in managing mobile privacy and security. By including the loved ones in CO-oPS network, individuals who are less knowledgeable about mobile privacy can receive oversight from those who have more knowledge. This app also has the potential to initiate open discussions regarding the app permission issues. From an older adults to a teen in families, CO-oPS can benefit a wide range of age groups and provide an interactive solution to keep everyone safe online and secure sensitive information.

\begin{acks}
We acknowledge the contributions of Nicholas Osaka, Anoosh Hari and Ricardo Mangandi, who developed the CO-oPS application. This research was supported by the U.S. National Science Foundation under grants CNS-1844881, CNS-1814068, CNS-1814110, and CNS-1814439. Any opinion, findings, and conclusions or recommendations expressed in this material are those of the authors and do not necessarily reflect the views of the U.S. National Science Foundation.
\end{acks}

\bibliographystyle{ACM-Reference-Format}
\bibliography{sample-base}


\begin{thebibliography}{22}


\ifx \showCODEN    \undefined \def \showCODEN     #1{\unskip}     \fi
\ifx \showDOI      \undefined \def \showDOI       #1{#1}\fi
\ifx \showISBNx    \undefined \def \showISBNx     #1{\unskip}     \fi
\ifx \showISBNxiii \undefined \def \showISBNxiii  #1{\unskip}     \fi
\ifx \showISSN     \undefined \def \showISSN      #1{\unskip}     \fi
\ifx \showLCCN     \undefined \def \showLCCN      #1{\unskip}     \fi
\ifx \shownote     \undefined \def \shownote      #1{#1}          \fi
\ifx \showarticletitle \undefined \def \showarticletitle #1{#1}   \fi
\ifx \showURL      \undefined \def \showURL       {\relax}        \fi
\providecommand\bibfield[2]{#2}
\providecommand\bibinfo[2]{#2}
\providecommand\natexlab[1]{#1}
\providecommand\showeprint[2][]{arXiv:#2}

\bibitem[nw_(2015)]%
        {nw_mobile_2015}
 \bibinfo{year}{2015}\natexlab{}.
\newblock \bibinfo{title}{Mobile apps, privacy and permissions: 5 key
  takeaways}.
\newblock
\newblock
\urldef\tempurl%
\url{https://www.pewresearch.org/fact-tank/2015/11/10/key-takeaways-mobile-apps/}
\showURL{%
\tempurl}


\bibitem[Agha et~al\mbox{.}(2021)]%
        {10.1007/978-3-030-77392-2_17}
\bibfield{author}{\bibinfo{person}{Zainab Agha}, \bibinfo{person}{Reza
  Ghaiumy~Anaraky}, \bibinfo{person}{Karla Badillo-Urquiola},
  \bibinfo{person}{Bridget McHugh}, {and} \bibinfo{person}{Pamela Wisniewski}.}
  \bibinfo{year}{2021}\natexlab{}.
\newblock \showarticletitle{`Just-in-Time' Parenting: A Two-Month Examination
  of the Bi-directional Influences Between Parental Mediation and Adolescent
  Online Risk Exposure}. In \bibinfo{booktitle}{\emph{HCI for Cybersecurity,
  Privacy and Trust}}, \bibfield{editor}{\bibinfo{person}{Abbas Moallem}}
  (Ed.). \bibinfo{publisher}{Springer International Publishing},
  \bibinfo{address}{Cham}, \bibinfo{pages}{261--280}.
\newblock
\showISBNx{978-3-030-77392-2}


\bibitem[Akter et~al\mbox{.}(2022)]%
        {10.1145/3512904}
\bibfield{author}{\bibinfo{person}{Mamtaj Akter}, \bibinfo{person}{Amy~J.
  Godfrey}, \bibinfo{person}{Jess Kropczynski}, \bibinfo{person}{Heather~R.
  Lipford}, {and} \bibinfo{person}{Pamela~J. Wisniewski}.}
  \bibinfo{year}{2022}\natexlab{}.
\newblock \showarticletitle{From Parental Control to Joint Family Oversight:
  Can Parents and Teens Manage Mobile Online Safety and Privacy as Equals?}
\newblock \bibinfo{journal}{\emph{Proc. ACM Hum.-Comput. Interact.}}
  \bibinfo{volume}{6}, \bibinfo{number}{CSCW1}, Article \bibinfo{articleno}{57}
  (\bibinfo{date}{apr} \bibinfo{year}{2022}), \bibinfo{numpages}{28}~pages.
\newblock
\urldef\tempurl%
\url{https://doi.org/10.1145/3512904}
\showDOI{\tempurl}


\bibitem[Aljallad et~al\mbox{.}(2019)]%
        {aljallad_designing_2019}
\bibfield{author}{\bibinfo{person}{Zaina Aljallad}, \bibinfo{person}{Wentao
  Guo}, \bibinfo{person}{Chhaya Chouhan}, \bibinfo{person}{Christy LaPerriere},
  \bibinfo{person}{Jess Kropczynski}, \bibinfo{person}{Pamela Wisnewski}, {and}
  \bibinfo{person}{Heather Lipford}.} \bibinfo{year}{2019}\natexlab{}.
\newblock \bibinfo{title}{Designing a {Mobile} {Application} to {Support}
  {Social} {Processes} for {Privacy} ({Journal} {Article}) {\textbar} {DOE}
  {PAGES}}.
\newblock
\newblock
\urldef\tempurl%
\url{https://par.nsf.gov/biblio/10097722}
\showURL{%
\tempurl}


\bibitem[Alsoubai et~al\mbox{.}(2022)]%
        {alsoubai_permission_2022}
\bibfield{author}{\bibinfo{person}{Ashwaq Alsoubai}, \bibinfo{person}{Reza
  Ghaiumy~Anaraky}, \bibinfo{person}{Yao Li}, \bibinfo{person}{Xinru Page},
  \bibinfo{person}{Bart Knijnenburg}, {and} \bibinfo{person}{Pamela~J.
  Wisniewski}.} \bibinfo{year}{2022}\natexlab{}.
\newblock \showarticletitle{Permission vs. App Limiters: Profiling Smartphone
  Users to Understand Differing Strategies for Mobile Privacy Management}. In
  \bibinfo{booktitle}{\emph{Proceedings of the 2022 CHI Conference on Human
  Factors in Computing Systems}} (New Orleans, LA, USA)
  \emph{(\bibinfo{series}{CHI '22})}. \bibinfo{publisher}{Association for
  Computing Machinery}, \bibinfo{address}{New York, NY, USA}, Article
  \bibinfo{articleno}{406}, \bibinfo{numpages}{18}~pages.
\newblock
\showISBNx{9781450391573}
\urldef\tempurl%
\url{https://doi.org/10.1145/3491102.3517652}
\showDOI{\tempurl}


\bibitem[Atkinson(2015)]%
        {noauthor_majority_2015}
\bibfield{author}{\bibinfo{person}{Michelle Atkinson}.}
  \bibinfo{year}{2015}\natexlab{}.
\newblock \bibinfo{title}{Majority of {U}.{S}. {Smartphone} {Owners} {Download}
  {Apps}}.
\newblock
\newblock
\urldef\tempurl%
\url{https://www.pewresearch.org/internet/2015/11/10/the-majority-of-smartphone-owners-download-apps/}
\showURL{%
\tempurl}


\bibitem[Badillo-Urquiola et~al\mbox{.}(2020)]%
        {badillo2020towards}
\bibfield{author}{\bibinfo{person}{Karla Badillo-Urquiola},
  \bibinfo{person}{Zainab Agha}, \bibinfo{person}{Mamtaj Akter}, {and}
  \bibinfo{person}{Pamela Wisniewski}.} \bibinfo{year}{2020}\natexlab{}.
\newblock \showarticletitle{Towards Assets-based Approaches for Adolescent
  Online Safety}. In \bibinfo{booktitle}{\emph{Badillo-Urquiola, Agha, Z.,
  Akter, K., Wisniewski, P.,(2020)“Towards Assets-Based Approaches for
  Adolescent Online Safety” Extended Abstract presented at the ACM Conference
  on Computer-Supported Cooperative Work Workshop on Operationalizing an
  Assets-Based Design of Technology,(CSCW 2020)}}.
\newblock


\bibitem[Calciati et~al\mbox{.}(2020)]%
        {calciati_automatically_2020}
\bibfield{author}{\bibinfo{person}{Paolo Calciati}, \bibinfo{person}{Konstantin
  Kuznetsov}, \bibinfo{person}{Alessandra Gorla}, {and}
  \bibinfo{person}{Andreas Zeller}.} \bibinfo{year}{2020}\natexlab{}.
\newblock \showarticletitle{Automatically {Granted} {Permissions} in {Android}
  apps: {An} {Empirical} {Study} on their {Prevalence} and on the {Potential}
  {Threats} for {Privacy}}. In \bibinfo{booktitle}{\emph{Proceedings of the
  17th {International} {Conference} on {Mining} {Software} {Repositories}}}
  \emph{(\bibinfo{series}{{MSR} '20})}. \bibinfo{publisher}{Association for
  Computing Machinery}, \bibinfo{address}{New York, NY, USA},
  \bibinfo{pages}{114--124}.
\newblock
\showISBNx{978-1-4503-7517-7}
\urldef\tempurl%
\url{https://doi.org/10.1145/3379597.3387469}
\showDOI{\tempurl}


\bibitem[Chouhan et~al\mbox{.}(2019)]%
        {chouhan_co-designing_2019}
\bibfield{author}{\bibinfo{person}{Chhaya Chouhan}, \bibinfo{person}{Christy~M.
  LaPerriere}, \bibinfo{person}{Zaina Aljallad}, \bibinfo{person}{Jess
  Kropczynski}, \bibinfo{person}{Heather Lipford}, {and}
  \bibinfo{person}{Pamela~J. Wisniewski}.} \bibinfo{year}{2019}\natexlab{}.
\newblock \showarticletitle{Co-designing for {Community} {Oversight}: {Helping}
  {People} {Make} {Privacy} and {Security} {Decisions} {Together}}.
\newblock \bibinfo{journal}{\emph{Proceedings of the ACM on Human-Computer
  Interaction}} \bibinfo{volume}{3}, \bibinfo{number}{CSCW}
  (\bibinfo{date}{Nov.} \bibinfo{year}{2019}), \bibinfo{pages}{1--31}.
\newblock
\showISSN{2573-0142, 2573-0142}
\urldef\tempurl%
\url{https://doi.org/10.1145/3359248}
\showDOI{\tempurl}


\bibitem[Das et~al\mbox{.}(2014)]%
        {das_effect_2014}
\bibfield{author}{\bibinfo{person}{Sauvik Das}, \bibinfo{person}{Tiffany
  Hyun-Jin Kim}, \bibinfo{person}{Laura~A. Dabbish}, {and}
  \bibinfo{person}{Jason~I. Hong}.} \bibinfo{year}{2014}\natexlab{}.
\newblock \showarticletitle{The effect of social influence on security
  sensitivity}. In \bibinfo{booktitle}{\emph{Proceedings of the {Tenth}
  {USENIX} {Conference} on {Usable} {Privacy} and {Security}}}
  \emph{(\bibinfo{series}{{SOUPS} '14})}. \bibinfo{publisher}{USENIX
  Association}, \bibinfo{address}{Menlo Park, CA}, \bibinfo{pages}{143--157}.
\newblock
\showISBNx{978-1-931971-13-3}


\bibitem[Davis and Davis(1989)]%
        {davis_perceived_1989}
\bibfield{author}{\bibinfo{person}{Fred Davis} {and} \bibinfo{person}{Fred
  Davis}.} \bibinfo{year}{1989}\natexlab{}.
\newblock \showarticletitle{Perceived Usefulness, Perceived Ease of Use, and
  User Acceptance of Information Technology}.
\newblock \bibinfo{journal}{\emph{" "}}  \bibinfo{volume}{13}
  (\bibinfo{year}{1989}), \bibinfo{pages}{319}.
\newblock
\urldef\tempurl%
\url{https://doi.org/10.2307/249008}
\showDOI{\tempurl}


\bibitem[Dourish et~al\mbox{.}(2004)]%
        {dourish_security_2004}
\bibfield{author}{\bibinfo{person}{Paul Dourish}, \bibinfo{person}{Rebecca~E.
  Grinter}, \bibinfo{person}{Jessica Delgado de~la Flor}, {and}
  \bibinfo{person}{Melissa Joseph}.} \bibinfo{year}{2004}\natexlab{}.
\newblock \showarticletitle{Security in the wild: user strategies for managing
  security as an everyday, practical problem}.
\newblock \bibinfo{journal}{\emph{Personal and Ubiquitous Computing}}
  \bibinfo{volume}{8}, \bibinfo{number}{6} (\bibinfo{date}{Nov.}
  \bibinfo{year}{2004}), \bibinfo{pages}{391--401}.
\newblock
\showISSN{1617-4917}
\urldef\tempurl%
\url{https://doi.org/10.1007/s00779-004-0308-5}
\showDOI{\tempurl}


\bibitem[Felt et~al\mbox{.}(2012)]%
        {felt_android_2012}
\bibfield{author}{\bibinfo{person}{Adrienne~Porter Felt},
  \bibinfo{person}{Elizabeth Ha}, \bibinfo{person}{Serge Egelman},
  \bibinfo{person}{Ariel Haney}, \bibinfo{person}{Erika Chin}, {and}
  \bibinfo{person}{David Wagner}.} \bibinfo{year}{2012}\natexlab{}.
\newblock \showarticletitle{Android permissions: user attention, comprehension,
  and behavior}. In \bibinfo{booktitle}{\emph{Proceedings of the {Eighth}
  {Symposium} on {Usable} {Privacy} and {Security}}}
  \emph{(\bibinfo{series}{{SOUPS} '12})}. \bibinfo{publisher}{Association for
  Computing Machinery}, \bibinfo{address}{New York, NY, USA},
  \bibinfo{pages}{1--14}.
\newblock
\showISBNx{978-1-4503-1532-6}
\urldef\tempurl%
\url{https://doi.org/10.1145/2335356.2335360}
\showDOI{\tempurl}


\bibitem[Ferreira et~al\mbox{.}(2015)]%
        {ferreira_securacy_2015}
\bibfield{author}{\bibinfo{person}{Denzil Ferreira}, \bibinfo{person}{Vassilis
  Kostakos}, \bibinfo{person}{Alastair~R. Beresford}, \bibinfo{person}{Janne
  Lindqvist}, {and} \bibinfo{person}{Anind~K. Dey}.}
  \bibinfo{year}{2015}\natexlab{}.
\newblock \showarticletitle{Securacy: an empirical investigation of {Android}
  applications' network usage, privacy and security}. In
  \bibinfo{booktitle}{\emph{Proceedings of the 8th {ACM} {Conference} on
  {Security} \& {Privacy} in {Wireless} and {Mobile} {Networks}}}
  \emph{(\bibinfo{series}{{WiSec} '15})}. \bibinfo{publisher}{Association for
  Computing Machinery}, \bibinfo{address}{New York, NY, USA},
  \bibinfo{pages}{1--11}.
\newblock
\showISBNx{978-1-4503-3623-9}
\urldef\tempurl%
\url{https://doi.org/10.1145/2766498.2766506}
\showDOI{\tempurl}


\bibitem[Ghaiumy~Anaraky et~al\mbox{.}(2021)]%
        {ghaiumy_disclose_2021}
\bibfield{author}{\bibinfo{person}{Reza Ghaiumy~Anaraky},
  \bibinfo{person}{Kaileigh~Angela Byrne}, \bibinfo{person}{Pamela~J.
  Wisniewski}, \bibinfo{person}{Xinru Page}, {and} \bibinfo{person}{Bart
  Knijnenburg}.} \bibinfo{year}{2021}\natexlab{}.
\newblock \showarticletitle{To Disclose or Not to Disclose: Examining the
  Privacy Decision-Making Processes of Older vs. Younger Adults}. In
  \bibinfo{booktitle}{\emph{Proceedings of the 2021 CHI Conference on Human
  Factors in Computing Systems}} (Yokohama, Japan) \emph{(\bibinfo{series}{CHI
  '21})}. \bibinfo{publisher}{Association for Computing Machinery},
  \bibinfo{address}{New York, NY, USA}, Article \bibinfo{articleno}{686},
  \bibinfo{numpages}{14}~pages.
\newblock
\showISBNx{9781450380966}
\urldef\tempurl%
\url{https://doi.org/10.1145/3411764.3445204}
\showDOI{\tempurl}


\bibitem[Kropczynski et~al\mbox{.}(2021)]%
        {10.1145/3479540}
\bibfield{author}{\bibinfo{person}{Jess Kropczynski}, \bibinfo{person}{Reza
  Ghaiumy~Anaraky}, \bibinfo{person}{Mamtaj Akter}, \bibinfo{person}{Amy~J.
  Godfrey}, \bibinfo{person}{Heather Lipford}, {and} \bibinfo{person}{Pamela~J.
  Wisniewski}.} \bibinfo{year}{2021}\natexlab{}.
\newblock \showarticletitle{Examining Collaborative Support for Privacy and
  Security in the Broader Context of Tech Caregiving}.
\newblock \bibinfo{journal}{\emph{Proc. ACM Hum.-Comput. Interact.}}
  \bibinfo{volume}{5}, \bibinfo{number}{CSCW2}, Article
  \bibinfo{articleno}{396} (\bibinfo{date}{oct} \bibinfo{year}{2021}),
  \bibinfo{numpages}{23}~pages.
\newblock
\urldef\tempurl%
\url{https://doi.org/10.1145/3479540}
\showDOI{\tempurl}


\bibitem[Mendel and Toch(2017)]%
        {mendel_susceptibility_2017}
\bibfield{author}{\bibinfo{person}{Tamir Mendel} {and} \bibinfo{person}{Eran
  Toch}.} \bibinfo{year}{2017}\natexlab{}.
\newblock \bibinfo{title}{Susceptibility to {Social} {Influence} of {Privacy}
  {Behaviors} {\textbar} {Proceedings} of the 2017 {ACM} {Conference} on
  {Computer} {Supported} {Cooperative} {Work} and {Social} {Computing}}.
\newblock
\newblock
\urldef\tempurl%
\url{https://dl.acm.org/doi/10.1145/2998181.2998323}
\showURL{%
\tempurl}


\bibitem[NW et~al\mbox{.}(2021)]%
        {nw_demographics_nodate}
\bibfield{author}{\bibinfo{person}{1615 L.~St NW}, \bibinfo{person}{Suite~800
  Washington}, {and} \bibinfo{person}{DC~20036 USA202-419-4300 {\textbar}
  Main202-857-8562 {\textbar} Fax202-419-4372 {\textbar}~Media Inquiries}.}
  \bibinfo{year}{2021}\natexlab{}.
\newblock \bibinfo{title}{Demographics of {Mobile} {Device} {Ownership} and
  {Adoption} in the {United} {States}}.
\newblock
\newblock
\urldef\tempurl%
\url{https://www.pewresearch.org/internet/fact-sheet/mobile/}
\showURL{%
\tempurl}


\bibitem[Rader and Wash(2015)]%
        {rader_identifying_2015}
\bibfield{author}{\bibinfo{person}{Emilee Rader} {and} \bibinfo{person}{Rick
  Wash}.} \bibinfo{year}{2015}\natexlab{}.
\newblock \showarticletitle{Identifying patterns in informal sources of
  security information}.
\newblock \bibinfo{journal}{\emph{Journal of Cybersecurity}}
  \bibinfo{volume}{1}, \bibinfo{number}{1} (\bibinfo{date}{Sept.}
  \bibinfo{year}{2015}), \bibinfo{pages}{121--144}.
\newblock
\showISSN{2057-2085}
\urldef\tempurl%
\url{https://doi.org/10.1093/cybsec/tyv008}
\showDOI{\tempurl}
\newblock
\shownote{Publisher: Oxford Academic}.


\bibitem[Reardon et~al\mbox{.}(2019)]%
        {reardon_50_2019}
\bibfield{author}{\bibinfo{person}{Joel Reardon}, \bibinfo{person}{Álvaro
  Feal}, \bibinfo{person}{Primal Wijesekera}, \bibinfo{person}{Amit Elazari~Bar
  On}, \bibinfo{person}{Narseo Vallina-Rodriguez}, {and} \bibinfo{person}{Serge
  Egelman}.} \bibinfo{year}{2019}\natexlab{}.
\newblock \showarticletitle{50 Ways to Leak Your Data: An Exploration of Apps'
  Circumvention of the Android Permissions System}. In
  \bibinfo{booktitle}{\emph{WINTER 2019, VOL. 44, NO. 4}} (2019).
  \bibinfo{publisher}{USENIX}, \bibinfo{address}{Boston, MA, United States},
  \bibinfo{pages}{603--620}.
\newblock
\showISBNx{978-1-939133-06-9}
\urldef\tempurl%
\url{https://www.usenix.org/conference/usenixsecurity19/presentation/reardon}
\showURL{%
\tempurl}


\bibitem[Schechter and Bonneau(2015)]%
        {schechter_learning_2015}
\bibfield{author}{\bibinfo{person}{Stuart Schechter} {and}
  \bibinfo{person}{Joseph Bonneau}.} \bibinfo{year}{2015}\natexlab{}.
\newblock \showarticletitle{Learning Assigned Secrets for Unlocking Mobile
  Devices}. In \bibinfo{booktitle}{\emph{" "}} (2015).
  \bibinfo{publisher}{USENIX}, \bibinfo{address}{" "},
  \bibinfo{pages}{277--295}.
\newblock
\showISBNx{978-1-931971-24-9}
\urldef\tempurl%
\url{https://www.usenix.org/conference/soups2015/proceedings/presentation/schechter}
\showURL{%
\tempurl}


\bibitem[Till and Densmore(2019)]%
        {till_characterization_2019}
\bibfield{author}{\bibinfo{person}{Sarina Till} {and} \bibinfo{person}{Melissa
  Densmore}.} \bibinfo{year}{2019}\natexlab{}.
\newblock \showarticletitle{A {Characterization} of {Digital} {Native}
  {Approaches} {To} {Mobile} {Privacy} and {Security}}. In
  \bibinfo{booktitle}{\emph{Proceedings of the {South} {African} {Institute} of
  {Computer} {Scientists} and {Information} {Technologists} 2019}}
  \emph{(\bibinfo{series}{{SAICSIT} '19})}. \bibinfo{publisher}{Association for
  Computing Machinery}, \bibinfo{address}{New York, NY, USA},
  \bibinfo{pages}{1--9}.
\newblock
\showISBNx{978-1-4503-7265-7}
\urldef\tempurl%
\url{https://doi.org/10.1145/3351108.3351131}
\showDOI{\tempurl}


\end{thebibliography}

\end{document}